# Optimizing Queries Using a Meta-level Database

**Christoph Koch**

Database and Artificial Intelligence Group
Technische Universität Wien, A-1040 Vienna, Austria
koch@dbai.tuwien.ac.at

**Abstract**

*Graph simulation* (using *graph schemata* or *data guides*) has been successfully proposed as a technique for adding structure to semistructured data. Design patterns for *description* (such as meta-classes and homomorphisms between schema layers), which are prominent in the object-oriented programming community, constitute a generalization of this graph simulation approach.

In this paper, we show description applicable to a wide range of data models that have some notion of object (-identity), and propose to turn it into a data model primitive much like, say, inheritance. We argue that such an extension fills a practical need in contemporary data management. Then, we present algebraic techniques for query optimization (using the notions of *described* and *description queries*). Finally, in the semistructured setting, we discuss the pruning of regular path queries (with nested conditions) using description meta-data. In this context, our notion of meta-data *extends* graph schemata and data guides by meta-level *values*, allowing to boost query performance and to reduce the redundancy of data.

## 1 Introduction

Research on the management of meta-data is an important issue in the database context. It is relevant to a wide variety of data management problems, including aspects of query optimization and physical storage management.

There are two major directions regarding meta-data in the data management context. The first one, which has received the greater share of attention, is the one of declaring the term meta-data equivalent to *database schemata* and providing tools and formalisms in which meta-data are closely integrated with the reasoning facilities. This view has resulted in a large body of research (e.g. higher-order logics like F-Logic [18] and HiLog [9], extended query languages such as SchemaSQL [22] and MD-SQL [28], and systems such as ConceptBase [16]).

An alternative view, that of meta-data as *data that describe data*, has received much coverage in the field of object-oriented software engineering [29, 5] – with work on meta-models and meta-classes, and design patterns with description semantics – but has seen only rudimentary formal treatment (e.g. [12]) and apparently no attention in the database arena. In this paper, we will focus on this second view of meta-data, as data explicitly stored in a database, featuring special description relationships with distinct "instance-level" data. The description semantics can be profitably applied to problems such as semantic query optimization [19, 14, 8, 13, 21, 24] and deciding the containment or equivalence of queries under meta-data interpreted as constraints (e.g., using the Chase [4, 25]), problems that are of wide practical interest.

Our argumentation builds on two simple concepts,

- the *description pattern* between classes and meta-classes (where each instance of a meta-class – that is, a meta-object – describes a certain category of objects, instances of classes), and

- the *homomorphism pattern* [29] between relationships of classes and relationships of meta-classes, entailing that for each link between objects that belongs to a relationship under a homomorphism on the instance level, a describing link on the meta-level must exist.

Although many of the terms used in this paper are borrowed from object-oriented data models, we argue that the results presented apply to many real-world databases using different data models (such as semantic relational and semistructured) as well.

**Example 1.1** Consider the case of a hypothetical database of a customer-centric car company. The company takes pride in providing each customer with a thoroughly customized product. This is achieved using a production management information system that provides flexibility on two levels.

- Firstly, the company follows an approach of offering a more fine-grained granularity of choices than a mere fixed number of car models in its product line. Rather, a number of *platforms* can be combined with a choice of engine models, dimensions of wheels and tires, body shapes, and interior design. Thus, the customer is offered a number of models of these constituent components, together with a compatibility matrix concerning their design specifications[1].

- The second degree of freedom is achieved by parameters that can be selected for each of the components, for instance the color of the car body or the materials to be used for furnishing the car's interior.

---

[1]Certainly, the customer needs help from a car dealer or software assistant on the Web when ordering his or her dream car.

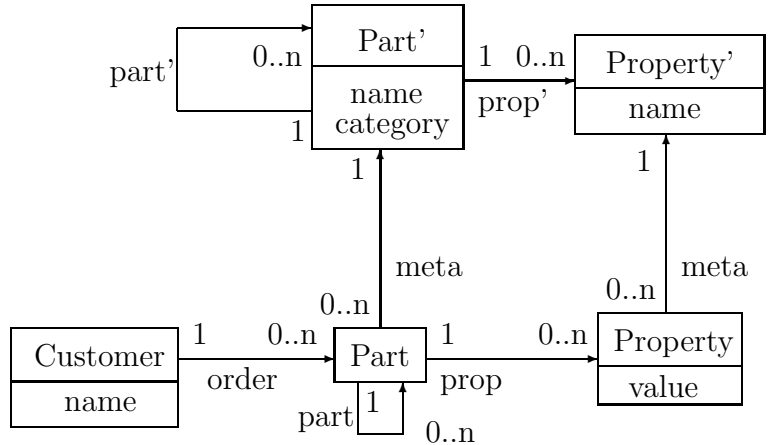

Figure 1: UML schema of Example 1.1.

The assembly status of each order can be tracked by the customer via the company's Web site. For traceability, information about assembly lines, operators, and parts from suppliers that are involved in the production of each order are recorded in the information system as assembly proceeds.

A simplified object-oriented database schema of our hypothetical information system is shown in Figure 1. This core schema covers the assembly trees of cars *as-built* (via the class *Part* and the association *part*), customer-order relationships, and the parameters chosen for the particular order (using the class *Property*). Car models and abstract components of cars *as-designed* are stored as *meta-objects* using the meta-classes *Part'* and *Property'*. For each object of the classes *Part* and *Property*, there is exactly one meta-object, reachable through its *meta* link. To put a maximum amount of abstract design-level information into this meta-level, the associations *Part'.part'* and *Part'.prop'* link meta-objects and *simulate* the associations *Part.part* and *Part.prop*, respectively, in the following natural manner: Whenever two objects are linked on the instance-level, their corresponding meta-objects are linked on the meta-level. For a simple database instance coherent with our schema, see Figure 2. This figure shows a customer order for a red sports utility vehicle (SUV), together with the meta-data describing the car model.

The layered schema design of Figure 1 has several strong points, including a neat separation of design-level and construction-level information, and the avoidance of redundancy. Note that description data are not automatically generated in some way by the system, but have been *carefully acquired* by our car company and are amongst its most valuable "treasures".

We address the problem of optimizing queries over such databases. Assume the following query (1):

select x from Part as x

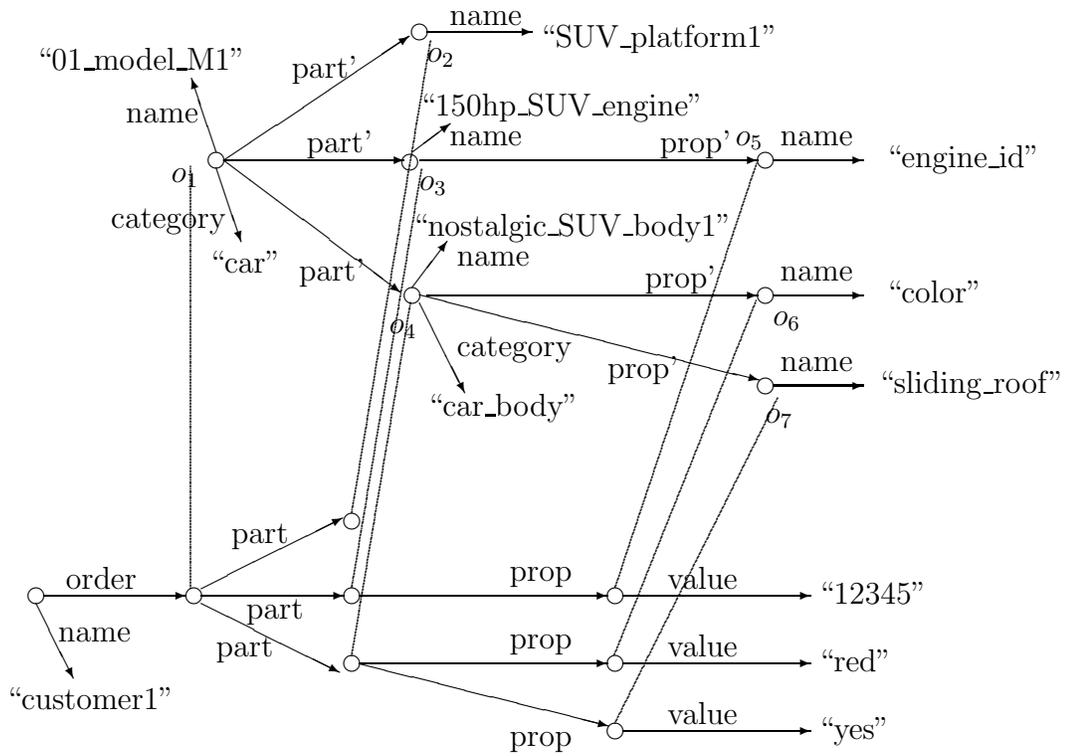

Figure 2: An example car order (bottom) and corresponding meta-data (above).

where exists x.part.prop as p
　　　where p.meta.name="color" and p.value = "red";

which asks for all parts with a direct subpart of red color. Assume now that we know of the integrity constraints that (a) only car bodies (i.e., objects of *Part'* with category="car_body") can have the property "color"

{"car_body"} = select body.category from Part' as body
　　　where exists body.prop' as p
　　　　　where p.name = "color";

and (b) parts that have direct subparts of category "car_body" are always of category "car", thus

{"car"} = select car.category from Part' as car
　　　where exists car.part' as body
　　　　　where body.category = "car_body";

It is now easy to see that under constraints (a) and (b), (1) is equivalent to

select car from Part as car
where car.meta.category = "car"
and exists car.part as body
　　　where body.meta.category = "car_body"
　　　and exists body.prop as p
　　　　　where p.meta.name="color" and p.value = "red";

Intuitively, what we have done to optimize (1) was to make use of the *restriction introduction* heuristics [19] of semantic query optimization. Alternatively to the use of integrity constraints, one may take advantage of the intuition that usually the amount of meta-data available is much smaller than the amount of instance-level data. Consider the following query (2).

select x from Part' as x
where exists x.part'.prop' as p
　　　where p.name="color";

It is easy to see that the meta-objects of the parts returned by our first query (1) must be contained in the result of (2). We call (2) the *description query* of (1). Since we may assume the meta-level database to be small, we may execute the meta-level-only query (2) and then use the result for optimizing (1). For instance, if our integrity constraints (a) and (b) hold, all objects in the result of (2) will be of category "car", and we can again optimize our input query by introducing this restriction. Moreover, if the result of (2) is empty, (1) is unsatisfiable.  □

The work presented in this paper requires that schemata and database applications are *designed* to make use of our description semantics; indeed, *we make description a data model primitive*. This is not an arbitrary decision, as these description semantics are very natural, and become increasingly widespread due to the popularity (and obvious merits) of object-oriented analysis and design and the influence of [29] on its community, for flexibility reasons as well as motivated by recent outspoken standardization efforts [27] following the approach of description advocated in this paper. Examples are large enterprise repositories and engineering databases [26, 10, 20, 30, 17], meta-data initiatives motivated by interoperability considerations [26], or database applications in which schemata need to be dynamically changeable and extensible (leaving aside schema evolution, however).

This leads us to semi-structured databases. In this context, the closely related mechanism of graph simulation is widely used for query optimization (e.g. [6, 1]) as well as a weak and flexible mechanism for typing objects (consider e.g. data guides [3] and graph schemata [7, 11]).

Also, note that the creation of dedicated structures for representing meta-data results often quite straightforwardly from a principled schema design process, independently from the underlying database paradigm or any special aspiration towards building a layered database.

The contributions of this paper are the following.

- We provide – to the best of our knowledge – the first treatment of the meta-data semantics used by the object-oriented software engineering community in a database context.

- We define a fragment of relational algebra which we call the *described queries* and which is associated to a class of *description queries* that can be obtained through a simple transformation. Given a described query, we can easily obtain its description query that will, when evaluated, return exactly all the meta-data that apply to the result of the described query. More exactly, there is a functional relationship between a tuple of a described query and the tuple that contains all the relevant meta-data.

- We show a few equivalences that allow to use results from a description query to optimize the described query – notably, each restriction obtained for a description query can directly be pushed down to the described query, and the emptiness of the result of the description query entails that the described query is empty, too. The rationale of this is that a meta-level database is usually by many orders of magnitude smaller than the corresponding instance-level database, while it usually contains much of the database values that are meaningful to users and are named in query conditions to restrict queries, and which are thus relevant for optimizing queries.

We go into the details of applying our results within the framework of the Chase in the case of conjunctive queries.

Our results are unostentatious yet elegant, permitting the application of optimizations analogous to those known for path queries (e.g. pruning using graph simulation [11]) to relational algebra-like query languages. [2]

- We approach the problem of optimizing regular path queries using a meta-level database in a semistructured data model. We present a new algorithm for optimizing path queries (with conditions) using our flavor of meta-data. In this context, we show that our meta-data semantics basically reduces to the well-known issue of graph simulation extended by description values. This extension allows for interesting new data management applications.

This paper is structured as follows. First we provide some preliminaries (Section 2). In Section 2.3, we discuss the meta-data semantics of the object-oriented software engineering community. Section 3 defines described and description queries and contains our results regarding query optimization in the framework of relational algebra. Section 4 contains our work on optimizing regular path queries with conditions in the semistructured database context. We conclude with a discussion of our work and its implications (Section 5).

## 2 Preliminaries

In this section, we first define an object-based data model which also generalizes from semantic relation-based data models. Next, to prepare the definition of our description semantics, we define a general notion of binary relationships, which is needed to be able to deal with, say, many-to-many relationships in data models that need to split such relationships into several relations. Finally, we provide definitions of our description semantics.

### 2.1 Data Model

The following sets of atomic elements are countably infinite and pairwise disjoint: a domain of *constants* $D = \{d_1, d_2, \ldots\}$, a set of *object identities* (oids) $\mathcal{O} = \{o_1, o_2, \ldots\}$, a set of *class names* $\{P_1, P_2, \ldots\}$, a set of *relation names* $\{R_1, R_2, \ldots\}$, and a set of *attribute names* $\{A_1, A_2, \ldots\}$. We use the notation $\langle A_1 : \cdots, \ldots, A_k : \cdots \rangle$ for tuples formed by $k \geq 1$ distinct attributes.

The set of *o-values* is the smallest set containing $D \cup \mathcal{O}$ s.t. if $v_1, \ldots, v_k$ (with $k > 0$) are o-values and $A_1, \ldots, A_k$ are attributes, then $\langle A_1 : v_1, \ldots, A_k : v_k \rangle$ and $\{v_1, \ldots, v_k\}$ are o-values as well. An *o-value assignment* for a finite set $\mathbf{R}$ of relation names is a function $\rho$ mapping each $R \in \mathbf{R}$ to a *finite* set of o-values. An *oid assignment* for a finite set $\mathbf{P}$ of class names is a function $\pi$ mapping each $P \in \mathbf{P}$ to a finite set of oids. $\pi$ is called *disjoint* if $\pi(P_1) \cap \pi(P_2) = \emptyset$ for all $P_1, P_2 \in \mathbf{P}$ with $P_1 \neq P_2$.

---
[2]The main reason why we use relational algebra (embedded into a semantic data model) is its simplicity; however, our results are easily translatable to algebraic object-oriented query languages.

The set of *type expressions*, $types(\mathbf{P})$, over a finite set of class names $\mathbf{P}$ is given by the following BNF syntax

$$\tau = D \mid P \mid \langle A_1 : \tau, \ldots, A_k : \tau \rangle \mid \{\tau\}$$

where $\tau$ is a type expression, $P \in \mathbf{P}$, and $k \geq 1$.

A *class hierarchy* $\langle \mathbf{P}, \mathbf{T}, \preceq \rangle$ is a tuple of a set of class names $\mathbf{P}$, a function $\mathbf{T} : \mathbf{P} \to types(\mathbf{P})$, and a partial order $\preceq$ over $\mathbf{P}$ for which we can define a *subtyping relationship* $\leq$ over $types(\mathbf{P})$ as follows. $\leq$ is the smallest partial order s.t.

1. $P_1 \preceq P_2$ implies $P_1 \leq P_2$

2. $\tau_1 \leq \tau_2$ implies $\{\tau_1\} \leq \{\tau_2\}$

3. if $\tau_{i,1} \leq \tau_{i,2}$ for each $i \in \{1, \ldots, n\}$, then

    $\langle A_1 : \tau_{1,1}, \ldots, A_n : \tau_{n,1}, B_1 : \tau_{1,3}, \ldots, B_m : \tau_{m,3} \rangle \leq \langle A_1 : \tau_{1,2}, \ldots, A_n : \tau_{n,2} \rangle$

$\langle \mathbf{P}, \mathbf{T}, \preceq \rangle$ is called *well-formed* if for each pair $P_1, P_2 \in \mathbf{P}$, $P_1 \preceq P_2$ implies $\mathbf{T}(P_1) \leq \mathbf{T}(P_2)$. For each class $P \in \mathbf{P}$, we can define the *extension* $\pi^*$ based on the disjoint extension $\pi$ as $\pi^*(P) = \{\pi(P_0) \mid P_0 \preceq P,\ P_0, P \in \mathbf{P}\}$.

The semantics of types is defined as follows. For each type expression $\tau$, its *disjoint interpretation* $[\![\tau]\!]_\pi$ is

1. $[\![D]\!]_\pi = D$

2. $[\![P]\!]_\pi = \pi^*(P)$ for each $P \in \mathbf{P}$

3. $[\![\langle A_1 : \tau_1, \ldots, A_k : \tau_k \rangle]\!]_\pi =$
   $\{\langle A_1 : v_1, \ldots, A_k : v_k \rangle \mid k \geq 1,\ v_i \in [\![\tau_i]\!]_\pi,\ i = 1, \ldots, k\}$

4. $[\![\{\tau\}]\!]_\pi = \{\{v_1, \ldots, v_k\} \mid v_i \in [\![\tau]\!]_\pi,\ i = 1, \ldots, k\}$

A *schema* $S$ is a tuple $\langle \mathbf{R}, \mathbf{P}, \mathbf{T}, \preceq \rangle$, where $\mathbf{R}$ is a finite set of relation names, $\mathbf{P}$ is a finite set of class names, $\mathbf{T}$ is a function from $\mathbf{R} \cup \mathbf{P}$ to $types(\mathbf{P})$, and $\langle \mathbf{P}, \mathbf{T}, \preceq \rangle$ is a well-formed class hierarchy. Of course, a relation is a set of tuples of atomic elements, and the types of relation names in $\mathbf{R}$ are defined accordingly.

An *instance* $\mathbf{I}$ of schema $\langle \mathbf{R}, \mathbf{P}, \mathbf{T}, \preceq \rangle$ is a triple $\langle \rho, \pi, \nu \rangle$, where $\rho$ is an o-value assignment for $\mathbf{R}$, $\pi$ is a disjoint oid assignment for $\mathbf{P}$, and $\nu$ is a (total) function from the set of oids in $\mathbf{I}$ to o-values, such that $\rho(R) \subseteq [\![\mathbf{T}(R)]\!]_\pi$ for each $R \in \mathbf{R}$, and $\nu(o) \in [\![\mathbf{T}(P)]\!]_\pi$ for each $P \in \mathbf{P}$ and each $o \in \pi(P)$. Each oid in $\mathbf{I}$ must belong to exactly one $\pi(P)$.

## 2.2 Semantic Relationships

Let $\tau_0, \ldots, \tau_n \in types(\mathbf{P})$. A *path expression* on type $\tau_0$ is an expression $\tau_0.A_1.\cdots.A_n$ such that for each $0 \leq i \leq n-1$, $t_i = \langle \ldots, A_{i+1} : \tau_{i+1}, \ldots \rangle$ or $t_i = \langle \ldots, A_{i+1} : \{\tau_{i+1}\}, \ldots \rangle$ where $t_i = T(\tau_i)$ if $\tau_i \in \mathbf{P}$ and $t_i = \tau_i$ otherwise.

The semantics of path expressions is given by their (binary) extension relations. For a path expression $\tau.A$ of length one we have the extension relation $[\![\tau.A]\!] = \{\langle v_1, v_2 \rangle \mid v_1 \in [\![\tau_1]\!] \text{ and } v_2 \in [\![\tau_2]\!] \text{ and } \dagger\}$

$$\dagger : \begin{cases} \nu(v_1) = \langle \ldots, A : v_2, \ldots \rangle & \ldots \; \tau_1 \in \mathbf{P}, T(\tau_1) = \langle \ldots, A : \tau_2, \ldots \rangle \\ \nu(v_1) = \langle \ldots, A : X, \ldots \rangle \text{ and } v_2 \in X & \ldots \; \tau_1 \in \mathbf{P}, T(\tau_1) = \langle \ldots, A : \{\tau_2\}, \ldots \rangle \\ v_1 = \langle \ldots, A : v_2, \ldots \rangle & \ldots \; \tau_1 \notin \mathbf{P}, T(\tau_1) = \langle \ldots, A : \tau_2, \ldots \rangle \\ v_1 = \langle \ldots, A : X, \ldots \rangle \text{ and } v_2 \in X & \ldots \; \tau_1 \notin \mathbf{P}, T(\tau_1) = \langle \ldots, A : \{\tau_2\}, \ldots \rangle \end{cases}$$

The extension relation $[\![\tau_0.A_1.\cdots.A_n]\!]$ for path expressions of length $n$ is defined as

$$[\![\tau_0.A_1.\cdots.A_n]\!] = [\![\tau_0.A_1]\!] \circ \ldots \circ [\![\tau_{n-1}.A_n]\!]$$

where $\circ$ denotes the composition of binary relations ($R_1 \circ R_2 = \{\langle v_1, v_3 \rangle \mid \langle v_1, v_2 \rangle \in R_1 \wedge \langle v_2, v_3 \rangle \in R_2\}$).

**Definition 2.1** A (binary) *relationship* $v$ between classes $P_1$ and $P_2$ may be of two forms.

1. The extension relation of a path expression $P_1.A_1.\cdots.A_n$ from $P_1$ to $P_2$, i.e. $v \subseteq [\![P_1]\!] \times [\![P_2]\!]$.

2. Conjunctive views

   $$v(O_1, O_2) \leftarrow R_1(\bar{X}_1) \wedge \ldots \wedge R_k(\bar{X}_k) \wedge v_1(O_{1,1}, O_{1,2}) \wedge \ldots \wedge v_m(O_{m,1}, O_{m,2}).$$

   where $R_1, \ldots, R_k \in \mathbf{R}$, $v_1, \ldots, v_m$ are binary relations of the former type (i.e., defined by path expressions), and $O_1, O_2$ appear in the $\bar{X}_i$ or $O_{i,j}$. We require the connectedness of the graph of the view body and that joins over oids are typed. Conjunctive views may contain constants from the domain of constants $D$ in relation columns typed $D$ in the $\bar{X}_i$.

□

A *simple* (binary) *relationship* between classes $P_1$ and $P_2$ is either a relation $R \in \mathbf{R}$ with $\mathbf{T}(R) = \langle A_1 : P_1, A_2 : P_2 \rangle$ or a path expression of length one.

**Example 2.2** Consider the schema of Example 1.1. We define four (simple) relationships

$$\vec{part} := Part.part \qquad \vec{part'} := Part'.part'$$
$$\vec{prop} := Part.prop \qquad \vec{prop'} := Part'.prop'$$

□

The need for binary relationships in their general form is justified by the intricacies that certain data models and schemata may offer.

**Example 2.3** Consider again our car domain from Example 1.1.

- Assume that the assembly tree of parts stores information on when and by whom a part was assembled. The type of the class *Part* could be

  $\mathbf{T}(Part) = \langle meta : Part',$
  $\phantom{\mathbf{T}(Part) = \langle} assembly : \{date : D,\ operator : D,\ part : Part\},$
  $\phantom{\mathbf{T}(Part) = \langle} prop : Property \rangle$

  The relationship $[\![Part.assembly.part]\!]$ covers the part-subpart semantics of the *part* association of Example 1.1.

- A directed edge-labeled graph can be represented using a single node class $P \in \mathbf{P}$ and a relation $R \in \mathbf{R}$ with $\mathbf{T}(R) = \langle P, D, P \rangle$. For instance, the fact that $o_2$ is a *part'* of $o_1$ (in Example 1.1) is evidenced by $\langle o_1, \text{part'}, o_2 \rangle \in R$. A binary relationship $\vec{part}'$ is thus defined as a view (here, in datalog notation)

  $$\vec{part}'(P_1, P_2) \leftarrow R(P_1, \text{part'}, P_2).$$

  $\square$

## 2.3 Description Semantics

**Definition 2.4** A *description relation* $P'$ desc $P$ between classes $P, P' \in \mathbf{P}$ satisfies the following requirements.

1. Its transitive closure is antisymmetric.

2. desc is closed with respect to $\preceq$: If $P'$ desc $P$, then $\{\langle P'_0, P_0 \rangle \mid P_0 \preceq P,\ P'_0 \preceq P'\} \subseteq$ desc .

3. desc is one-to-one up to inheritance: If $P'_1$ desc $P$ and $P'_2$ desc $P$, then $P'_1 \preceq P'_2$ or $P'_2 \preceq P'_1$. If $P'$ desc $P_1$ and $P'$ desc $P_2$, then $P_1 \preceq P_2$ or $P_2 \preceq P_1$.

We call a class $P$ *described* iff there is a class $P'$ s.t. $P'$ desc $P$. $P'$ is called (the) *meta-class* of $P$. We define a function $\mu$ which maps each oid $o \in [\![P]\!]_\pi$ to an oid $\mu(o) \in [\![P']\!]_\pi$, for all $P, P' \in \mathbf{P}$ s.t. $P'$ desc $P$. $\square$

Next we adapt the notion of *homomorphisms* in object-oriented schemata of [29].

**Definition 2.5** A *homomorphism* relation is a binary relation over binary relationships. Let $V'$ hom $V$ with $V' \subseteq [\![P_1']\!]_\pi \times [\![P_2']\!]_\pi$ and $V \subseteq [\![P_1]\!]_\pi \times [\![P_2]\!]_\pi$. Then,

1. $P_1'$ desc $P_1$ and $P_2'$ desc $P_2$ and

2. The transitive closure of hom is antisymmetric.

3. hom is one-to-one: If $V'$ hom $V_1$ and $V'$ hom $V_2$, then $V_1 = V_2$. If $V_1'$ hom $V$ and $V_2'$ hom $V$, then $V_1' = V_2'$.

4. $\langle o_1, o_2 \rangle \in V$ implies $\langle \mu(o_1), \mu(o_2) \rangle \in V'$ for all $o_1 \in [\![P_1]\!]_\pi$ and $o_2 \in [\![P_2]\!]_\pi$.

□

Relationships related by homomorphisms entail a natural layering of schemata (which may be partial, however). Note that both desc and hom can be defined to be transitive, which is practical if more than one layer of description exists. However, then, the definition of $\mu$ becomes somewhat problematic.

**Definition 2.6** A *schema with description* is a tuple $\langle \mathbf{R}, \mathbf{P}, \mathbf{T}, \preceq, \text{desc}, \text{hom} \rangle$, where $\langle \mathbf{R}, \mathbf{P}, \mathbf{T}, \preceq \rangle$ is a schema, desc is a description relation on $\mathbf{P}$, and hom is a homomorphism relation over $\mathbf{R}$, $\mathbf{P}$, and desc. An *instance with description* is a tuple $\langle \rho, \pi, \nu, \mu \rangle$, where $\langle \rho, \pi, \nu \rangle$ is an instance and $\mu$ is defined as above. □

**Example 2.7** Consider the schema of Example 1.1 and the relationships of Example 2.2. For the database schema $S = \langle \mathbf{R} = \emptyset, \mathbf{P}, \mathbf{T}, \preceq, \text{desc}, \text{hom} \rangle$ of Example 1.1, we have

$$
\begin{aligned}
\mathbf{P} &= \{Part, Property, Customer, Part', Property'\} \\
\mathbf{T}(Part) &= \langle part : \{Part\}, prop : \{Property\}, meta : Part \rangle \\
\mathbf{T}(Part') &= \langle name : D, category : D, part : \{Part'\}, prop : \{Property'\} \rangle \\
\mathbf{T}(Property) &= \langle meta : \{Property'\}, value : D \rangle \\
\mathbf{T}(Property') &= \langle name : D \rangle \\
\text{desc} &= \{\langle Part', Part \rangle, \langle Property', Property \rangle\} \\
\text{hom} &= \{\langle \vec{part'}, \vec{part} \rangle, \langle \vec{prop'}, \vec{prop} \rangle\}
\end{aligned}
$$

Inheritance is not used, thus $\preceq$ is the identity relation on $\mathbf{P}$.

The instance of Example 1.1 is shown in Figure 2. For example,

$$\nu(o_1) = \langle name : \text{``01\_model\_M1''}, category : \text{``car''}, part' : \{o_2, o_3, o_4\}, prop' : \emptyset \rangle$$

and $\rho = \emptyset$, $[\![Part']\!]_\pi = \{o_1, o_2, o_3, o_4\}$, $[\![Property']\!]_\pi = \{o_5, o_6, o_7\}$. As the UML data model lacks description semantics, we have represented description relationships by mandatory associations "meta" from classes to meta-classes. We have $\mu(o) = o.meta$ for all objects $o$ of described classes. □

# 3 Described Queries

In this section, we present some of the main ideas of this paper. These ideas – centering around the notions of described and description queries – are not so much dependent on a particular data model (a notion of object or entity, however, is required), and even less so on a particular query language. To make our point as simple and intuitive as possible, we use a query language that is based on well-known (positive) relational algebra. However, we dare a few modifications.

- Queries are based on classes and binary relationships (as presented earlier). Let $P$ be a class. Then, $P$ is a query with the semantics $\{\langle o \rangle \mid o \in [\![P]\!]_\pi\}$. (Thus, we obtain a set of tuples of objects rather than a set of objects. This is a shortcut to simplify the subsequent presentation.)

- All joins are of the form $Q_1 \bowtie R \bowtie Q_2$ (abbreviated $Q_1 \bowtie_R Q_2$), where $R$ is a binary relationship, or $R \bowtie Q$ (abbreviated $\bowtie_R Q$), where *both* columns of $R$ are joined with $Q$. [3]

- Queries are typed in compliance with **P** and $\preceq$ (i.e., only columns of the same type $\tau = P \in \mathcal{P}$ may be joined).

- The relational selection operation $\sigma$ tests attributes typed $D$ of the objects occurring in query tuples rather than the objects themselves (see Example 3.1).

- We define a new relational selection operator $\sigma'$ supporting our intuition that the attributes of a meta-class are in some sense also owned by the class described by it. The semantics of the operation is

$$\sigma'_\gamma(P) = \{\langle o \rangle \mid o \in [\![P]\!]_\pi,\ \langle \mu(o) \rangle \in \sigma_\gamma(P')\},$$

where $P'$ desc $P$, and $\gamma$ is a boolean selection condition (using $\wedge, \vee, \neg$) over attributes of $P'$.

By $\$i$, we denote the $i$-th column of a query.

**Example 3.1** Consider again query (1) of our running example. In our query language, it is formulated as $Q_1 = \pi_{\$1}(\sigma_{\$3.\text{value} = \text{``red''}}(Q'_1))$ with

$$Q'_1 = (Part \bowtie_{p\vec{a}rt} Part) \bowtie_{p\vec{r}op(\$2,\$1)} \sigma'_{\text{name} = \text{``color''}}(Property)$$

The type of $Q_1$ is $\langle Part \rangle$ and $\mathbf{T}(Q'_1) = \langle Part, Part, Property \rangle$. □

---
[3]Thus, such joins also cover path expressions of object-oriented query languages given appropriately defined relationships.

**Definition 3.2** Let $\langle \mathbf{R}, \mathbf{P}, \mathbf{T}, \preceq, \text{desc}, \text{hom} \rangle$ be a layered database schema. The set of *described queries* (DQ) is the smallest set that satisfies

1. If $P'$ desc $P$, then $P$ is a DQ.

2. If $Q_1$ and $Q_2$ are described queries with the types $\mathbf{T}(Q_1) = \langle \ldots, A_i : P_1, \ldots \rangle$ and $\mathbf{T}(Q_2) = \langle \ldots, B_j : P_2, \ldots \rangle$, $R \subseteq [\![P_1]\!]_\pi \times [\![P_2]\!]_\pi$, and $R'$ hom $R$, then $(Q_1 \bowtie_{R(A_i, B_j)} Q_2)$ is a DQ.

3. If $Q$ is a DQ, $\mathbf{T}(Q) = \langle \ldots, A_i : P_1, \ldots, A_j : P_2, \ldots \rangle$, $R \subseteq [\![P_1]\!]_\pi \times [\![P_2]\!]_\pi$, and $R'$ hom $R$, then $(\bowtie_{R(A_i, A_j)} Q)$ is a DQ.

4. If $Q$ is a DQ and $\gamma$ is a selection condition over a class in $M(Q)$, then $\sigma'_\gamma(Q)$ is a DQ.

5. If $Q$ is a DQ with $\mathbf{T}(Q) = \langle A_1 : P_1, \ldots, A_n : P_n \rangle$, then $\pi_{A_{i_1}, \ldots, A_{i_m}}(Q)$ with $i_1, \ldots, i_m \in \{1, \ldots, n\}$ is a DQ.

6. If $Q_1$ and $Q_2$ are described queries with $\mathbf{T}(Q_1) = \mathbf{T}(Q_2)$, then $(Q_1 \cup Q_2)$ and $(Q_1 \cap Q_2)$ are described queries. □

The function $M$ translates (rewrites) a described query into its corresponding query on the meta-level:

**Definition 3.3** For each described query $Q$, the function $M$ maps $Q$ to its so-called corresponding *description query*. $M$ is defined as

$$\begin{aligned} M(P) &= P' \\ M(Q_1 \bowtie_R Q_2) &= M(Q_1) \bowtie_{R'} M(Q_2) \\ M(\bowtie_R (Q)) &= \bowtie_{R'} (M(Q)) \\ M(\sigma'_\gamma(Q)) &= \sigma_\gamma(M(Q)) \\ M(\pi_{A_{i_1}, \ldots, A_{i_m}}(Q_0)) &= \pi_{A_{i_1}, \ldots, A_{i_m}}(M(Q_0)) \\ M(Q_1 \cap Q_2) &= M(Q_1) \cap M(Q_2) \\ M(Q_1 \cup Q_2) &= M(Q_1) \cup M(Q_2) \end{aligned}$$

if $P'$ desc $P$, $R'$ hom $R$, and $\mathbf{T}(Q_0) = \langle A_1 : P_1, \ldots, A_n : P_n \rangle$, $i_1, \ldots, i_m \in \{1, \ldots, n\}$, $P_1, \ldots, P_n \in \mathbf{P}$; for $Q_1 \cap Q_2$ and $Q_1 \cup Q_2$: if $\mathbf{T}(Q_1) = \mathbf{T}(Q_2)$. □

**Example 3.4** Query $Q'_1$ of the previous example is a described query. We have

$$M(Q'_1) = (Part' \bowtie_{\vec{part'}} Part') \bowtie_{\vec{prop'}(\$2,\$1)} \sigma_{\text{name} = \text{``color''}}(Property')$$

□

It is easy to see that $M$ is reversible, i.e., that we can use $M^{-1}$ to translate queries "down" one meta-level.

We overload $\mu$ as follows. Let $Q$ be a described query and $\bar{t} = \langle o_1, \ldots, o_n \rangle$ be a tuple of $Q$. Then, $\mu.\bar{t} = \langle \mu(o_1), \ldots, \mu(o_n) \rangle$ (i.e., $\mu.\bar{t}$ denotes the element-wise application of $\mu$ to $\bar{t}$) and $\mu(Q) = \{\mu.\bar{t} \mid \bar{t} \in Q\}$. We can define an operator $\ltimes_\mu$ analogous to the relational semijoin operator s.t. $Q \ltimes_\mu Q' = \{\bar{t} \in Q \mid \mu.\bar{t} \in Q'\}$ (where $Q$ must be a described query). Because of the linearity of $\mu$, we have $\mu(Q_1 \times Q_2) = \mu(Q_1) \times \mu(Q_2)$ and $\mu(\pi_S(Q)) = \pi_S(\mu(Q))$ (where $S$ is any set of columns to be projected).

The important property of the description queries is that for each tuple in a described query ($\bar{t} \in Q$) there is a tuple $\bar{t}' \in M(Q)$ that contains all of its meta-data. This leads us to the following theorem.

**Theorem 3.5** Let $Q$ be a described query. For all $\bar{t} \in Q$, $\mu.\bar{t} \in M(Q)$. □

We can also write this as $\mu(Q) \subseteq M(Q)$.

**Proof Sketch** It is easy to show $\mu(Q) \subseteq M(Q)$ by induction in the size of the described query. The induction starts with atomic queries of the form $P \in \mathbf{P}$: Let $P'$ desc $P$. By definition of $\mu$, we have $\mu(P) \subseteq P' = M(P)$.

For the remaining steps, we assume that $\mu(Q) \subseteq M(Q)$ holds for queries $Q$ ($Q_1$, $Q_2$, ...) of the next smaller size.

1. We split joins into the computation of the cartesian product (where appropriate) and a selection operation $\bowtie_R$. Let $M(Q_1 \times Q_2) = M(Q_1) \times M(Q_2)$. It is clear that $\mu(Q_1 \times Q_2) = \mu(Q_1) \times \mu(Q_2) \subseteq M(Q_1) \times M(Q_2)$ because $\mu(Q_i) \subseteq M(Q_i)$.

   $\mu(\bowtie_R (Q)) \subseteq \bowtie_{R'} (M(Q))$ if $R'$ hom $R$ by the definition of the homomorphism semantics (4): $\langle o_1, o_2 \rangle \in R$ implies $\mu(\langle o_1, o_2 \rangle) \in R'$.

2. Let $\gamma$ be over class $P$ only in $Q$. We can push the selection $\sigma'_\gamma(Q)$ down the syntax tree of the query until we obtain a subexpression $\sigma'_\gamma(P)$. However, by definition, $\sigma'_\gamma(P) = P \ltimes_\mu \sigma_\gamma(P')$. Since $\mu(\sigma'_\gamma(P)) = \mu(P \ltimes_\mu \sigma_\gamma(M(P))) = \sigma_\gamma(M(P)) = M(\sigma'_\gamma(P))$, our claim holds.

3. Projection: our claim immediately follows from the premise $\mu(Q) \subseteq M(Q)$ and the linearity of $\mu$.

4. Union: $\mu(Q_1 \cup Q_2) = \mu(Q_1) \cup \mu(Q_2)$. Because of $\mu(Q_i) \subseteq M(Q_i)$, $\mu(Q_1) \cup \mu(Q_2) \subseteq M(Q_1) \cup M(Q_2)$.

5. Intersection: as for union. □

Now, for described queries $Q$ and $\sigma'_\gamma(Q)$, we have $Q \equiv Q \ltimes_\mu M(Q)$ and $\sigma'_\gamma(Q) \equiv Q \ltimes_\mu \sigma_\gamma(M(Q))$, respectively.

**Corollary 3.6** For a described query $Q$,

1. if $M(Q)$ is unsatisfiable or empty, $Q$ is unsatisfiable, and

2. if $M(Q) \equiv \sigma_\gamma(M(Q))$, then $Q \equiv \sigma'_\gamma(Q)$.

□

Note that Theorem 3.5 does not hold if we extend described queries by $Q_1 \setminus Q_2$ (for described queries $Q_1$ and $Q_2$) with $M(Q_1 \setminus Q_2) = M(Q_1) \setminus M(Q_2)$.

**Example 3.7** Let $P'_1$ desc $P_1$, $P'_2$ desc $P_2$, $R'$ hom $R$, $\pi(P'_1) = \{o'_1\}$, $\pi(P'_2) = \{o'_2\}$, $\pi(P_1) = \{o_{1,1}, o_{1,2}\}$, $\pi(P_2) = \{o_2\}$, $R = \{\langle o_{1,2}, o_2 \rangle\}$, $R' = \{\langle o'_1, o'_2 \rangle\}$, and $Q = P_1 \setminus \pi_{\$1}(P_1 \bowtie_R P_2)$. We have $Q = \{\langle o_{1,1} \rangle\}$ and $M(Q) = P'_1 \setminus \pi_{\$1}(P'_1 \bowtie_{R'} P'_2) = \emptyset$; thus, $\mu(Q) \not\subseteq M(Q)$. □

Let us now consider the semantic query optimization problem [14, 19, 8], where we basically want to use semantics in the form of *integrity constraints* to optimize queries.

A known problem with integrity constraints encoding data semantics (differently from dependencies that encode schema semantics such as foreign key constraints) is that in real-world databases, such constraints are rarely available, because providing them requires a fair amount of work and understanding of database issues from users. For this reason, there has been work on automatically deriving (mining) useful integrity constraints [32]. This of course requires to make a closed-world assumption for the data; integrity constraints have to be changed when updates to the database occur. Now, the application of such techniques to meta-level databases has special advantages; particularly, meta-data change less frequently than instance-level data and the size of meta-data may be by many orders of magnitude smaller than the corresponding instance-level data. Still, most of the conditions in practical queries are actually based on descriptive values[4] that are likely to be represented as meta-data if appropriate means for description (as they are proposed in this paper) are available.

Our results motivate a number of ways to optimize queries.

- Given a query $Q$, we can find subexpressions of $Q$ which are described queries. For those, we compute the corresponding description queries and apply semantic query optimization techniques to them, notably the *restriction introduction* and *unsatisfiability detection* heuristics [14, 19, 8] (using Corollary 3.6). The results found are then applied to the input query $Q$.

- Alternatively, one can simply execute description queries and apply the results to original queries. The intuition here is that meta-data tend to be much smaller that instance-level data, which makes such an approach practically feasible.

---

[4]We may expect users' memories to be focussed on the essentials.

- A third alternative is to translate integrity constraints derived for the meta-level down to the instance level, basically by using the inverse of the $M$ function. These constraints can subsequently be used to optimize queries using a conventional optimizer.

Note that the first two optimization approaches were explained earlier in Example 1.1. The algebraic approach followed in this section allows to easily implement our method in many algebraic query optimizers.

**Conjunctive Queries**

Let us briefly consider the case of conjunctive described queries, which are described queries without the union and set difference operations and where selection conditions do not use the conjuncts $\vee$ or $\neg$. We use a standard logical notation for conjunctive queries.

For conjunctive queries, we want to be able to incorporate our notions of described and description queries into the successful Chase framework [4, 25, 2]. This is to be able to apply previous results (see e.g. [2]) on optimality for many classes of integrity constraints.

Indeed, this goal is easily achieved by merging[5] a described query $Q$ with its description query and optimize the combined query $Q \bowtie_\mu M(Q)$. To such a combined query, we can then apply the Chase procedure with numerous classes of integrity constraints. A natural class of constraints well-suited for semantic query optimization (and the automatal inferral of constraints from data) is the *implication integrity constraints* [35, 31]. Two examples of such constraints are (a) and (b) in Example 1.1. The problems of optimizing conjunctive queries using implication integrity constraints and their incorporation into the Chase are well understood [35, 31]. Informally, all we need to do is to consider the atoms of the query body as a frozen database and to interpret a set of implication integrity constraints like a datalog program on this database.

**Example 3.8** Consider the query $Q'_1$ of our running example. $Q := Q'_1 \bowtie_\mu M(Q'_1)$ is a conjunctive query

$$Q(\bar{X}) \leftarrow Part(X_1), \vec{part}(X_1, X_2), Part(X_2), \vec{prop}(X_2, X_3), Property(X_3),$$
$$Part'(X'_1), \vec{part}'(X'_1, X'_2), Part'(X'_2), \vec{prop}'(X'_2, X'_3), Property(X'_3),$$
$$\mu(X_1, X'_1), \mu(X_2, X'_2), \mu(X_3, X'_3), (X'_3.name = \text{``color''}).$$

---
[5]To determine described subexpressions of an arbitrary conjunctive query $Q$, we first need to rewrite $Q$ into a query that uses the relationships (defined as conjunctive views in Definition 2.1) for which we have defined homomorphisms. This is the problem of rewriting queries using views, which is well understood (e.g. [23]).

The integrity constraints (a) and (b) of Example 1.1 are implication integrity constraints of the form

$$(X.category = \text{``car\_body''}) \leftarrow Part'(X), \vec{prop'}(X,Y), Property'(Y),$$
$$(Y.name = \text{``color''}).$$
$$(X.category = \text{``car''}) \leftarrow Part'(X), \vec{part'}(X,Y), Part'(Y),$$
$$(Y.category = \text{``car\_body''}).$$

As pointed in Example 1.1, we can optimize $Q$ by first introducing the restriction $(X'_2.category = \text{``car\_body''})$ using (a) and then $(X'_1.category = \text{``car''})$ using (b). □

By considering $Q \ltimes_\mu M(Q)$ rather than $Q$, we have introduced some redundancies that may be eliminated at the end of the Chase (as we usually understand by global optimality (also) a minimal number of joins) by first removing all meta-relationships simulating instance-level relationships present in the query and then removing those occurrences of meta-classes that are not made necessary by $\sigma'$-selections.

Note, however, that in practice seemingly redundant meta-level relationships may help to reduce query execution costs. If, say, $Q = P_1 \bowtie_R P_2$ and $M(Q)$ is much easier to evaluate than $Q$, the equivalent query $(P_1 \ltimes_\mu M(Q)) \bowtie_R (P_2 \ltimes_\mu M(Q))$ may be faster to execute than $Q$.

## 4 Optimizing Path Queries

In this section, we apply the concepts introduced so far to a semistructured database setting, and assume data to be schema-less. To keep up with the notions used so far, **P** consists of two classes, namely instance-level nodes and meta-level nodes. There are, however, several relationships, distinguished by labels. In order not to need to explicitly represent the hom relation, we make the assumption that relationships on the instance and meta-levels are under a homomorphism iff their names are the *same*. Each instance-level node or edge is described.

In the following, let $T$ be a set of tag names, $A$ be a set of attribute names, and the language $\mathcal{L}_A$ be defined as the set of all attribute assignments $a = s$, where $a \in A$ and $s$ is a string.

**Definition 4.1** A *graph database* is a node-labeled rooted graph $G = \langle V, r, lab, E \rangle$, where $r$ is the only node in $V$ whose in-degree is 0 and $lab : V \to T \times 2^{\mathcal{L}_A}$ is the labeling function. □

We use the abbreviation *tag* s.t. $tag(v) = t$ iff $\exists X : lab(v) = \langle t, X \rangle$. Given a graph database $I = \langle V_I, r_I, lab_I, E_I \rangle$, a graph database $M = \langle V_M, r_M, lab_M, E_M \rangle$ is a *meta-level database* for $I$ if we can define a function $\mu : V_I \to V_M$ s.t.

```
<db description=true>
<part id="01_model_M1" category="car">
   <part id="suv_platform1" />

   <part id="150hp_suv_engine1"><prop name="engine_id" /></part>

   <part id="nostalgic_suv_body1" category="car_body">
      <prop name = "color" />
      <prop name = "sliding_roof" />
   </part>
</part>
</db>
```

Figure 3: Meta-data of Example 1.1 as an XML file.

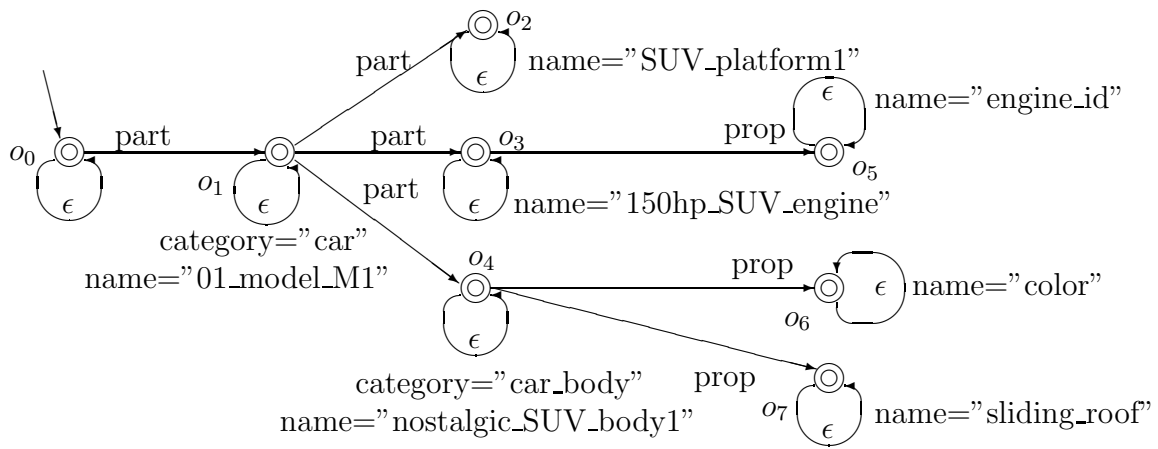

Figure 4: Meta-data FSA for Example 1.1.

1. for each $v \in V_I$, $tag_I(v) = tag_M(\mu(v))$ and
2. for each pair $v_1, v_2 \in V_I$, $\langle v_1, v_2 \rangle \in E_I \Rightarrow \langle \mu(v_1), \mu(v_2) \rangle \in E_M$.

We consider the following query language.

**Definition 4.2** (Query language). The abstract syntax of regular path queries with nested conditions is defined by the grammar

| | |
|---|---|
| *start*: | *path* |
| *path*: | tag \| *path* '.' *path* \| *path* '\|' *path* \| '(' *path* ')*' \| *path* '[' *conds* ']' |
| *conds*: | *cond* \| *cond* 'and' *conds* |
| *cond*: | *path* \| attr '=' string |

where "attr" is a set of attribute names, "tag" the set of HTML tag names, and "string" the set of strings. ('.' and '|' are associative, respectively.) We denote this query language by $\mathcal{L}_Q$.

The semantics of our query language is the one of classical regular path queries extended by a fragment of XPath [33] conditions. Notably, a query as a condition is true iff it returns a nonempty result w.r.t. the current context node.

Given the graph database $G = \langle V, r, lab, E \rangle$. Let $\pi, \pi_1, \pi_2 \in \mathcal{L}_Q$ denote queries in our language and $\gamma, \gamma_1, \gamma_2 \in \mathcal{L}_\gamma$ denote conditions. We define the semantic function $\mathbb{E}: \mathcal{L}_Q \times V \to 2^V$ for query expressions and $\mathbb{C}: \mathcal{L}_\gamma \times V \to \{\text{true}, \text{false}\}$ for conditions.

$$\begin{aligned}
\mathbb{E}[\![t]\!]n &= \{n \mid lab(n) = \langle t, X \rangle\} \\
\mathbb{E}[\![\pi_1.\pi_2]\!]n &= \{n''' \mid n' \in \mathbb{E}[\![\pi_1]\!]n \wedge n''' \in \mathbb{E}[\![\pi_2]\!]n'' \wedge E(n'', n''')\} \\
\mathbb{E}[\![\pi_1 \mid \pi_2]\!]n &= \mathbb{E}[\![\pi_1]\!]n \cup \mathbb{E}[\![\pi_2]\!]n \\
\mathbb{E}[\![\pi^*]\!]n &= \{n' \mid \langle n, n' \rangle \in R^*_\pi\} \\
\mathbb{E}[\![\pi[\gamma]]\!]n &= \{n' \mid n' \in \mathbb{E}[\![\pi]\!] \wedge \mathbb{C}[\![\gamma]\!]n' = \text{true}\} \\
\mathbb{C}[\![\gamma_1 \wedge \gamma_2]\!]n &= \mathbb{C}[\![\gamma_1]\!]n \wedge \mathbb{C}[\![\gamma_2]\!]n \\
\mathbb{C}[\![\pi]\!]n &= |\mathbb{E}[\![\pi]\!]n| > 0 \\
\mathbb{C}[\![a = s]\!]n &= \begin{cases} \text{true} & \ldots \; lab(n) = \langle t, X \rangle \wedge \langle a, s \rangle \in X \\ \text{false} & \ldots \; \text{otherwise} \end{cases}
\end{aligned}$$

where $R_\pi = \{\langle n, n' \rangle \mid n' \in \mathbb{E}[\![\pi]\!]n\}$ and $R^*_\pi$ is the reflexive and transitive closure of the relation $R_\pi$. The semantics of query $Q$ is $\mathbb{E}[\![Q]\!]r$. □

Regular path queries are a powerful and theoretically elegant formalism. The extension by nested conditions is required to be able to make use of (meta-level) values that are important to our approach. To improve readability, we will underline queries and attribute selection expressions in the remainder of this paper.

**Definition 4.3** Given a graph database $G = \langle V, r, lab, E \rangle$, its (nondeterministic) finite state automaton (FSA) $\mathcal{A}$ is defined as $\mathcal{A} = \langle Q = V, s = r, \delta, F = Q \rangle$, where

$$\begin{aligned}
\delta &= \{\langle v_1, \langle t, \emptyset \rangle, v_2 \rangle \mid \langle v_1, v_2 \rangle \in E \wedge lab(v_2) = \langle t, C \rangle\} \\
&\cup \{\langle v, \langle \epsilon, C \rangle, v \rangle \mid v \in V, \ lab(v) = \langle t, C \rangle\}
\end{aligned}$$

We denote $\mathcal{A}$ by FSA($G$). $\square$

**Algorithm 4.4** Given a query $Q \in \mathcal{L}_Q$, its regular expression $[\![Q]\!]$ is computed using the productions

$$[\![\pi_1.\pi_2]\!] \Rightarrow [\![\pi_1]\!].[\![\pi_2]\!] \qquad [\![\pi_1 \mid \pi_2]\!] \Rightarrow [\![\pi_1]\!] \mid [\![\pi_2]\!] \qquad [\![\pi^*]\!] \Rightarrow [\![\pi]\!]^* \qquad [\![(\pi)]\!] \Rightarrow ([\![\pi]\!])$$

$$[\![t]\!] \Rightarrow \langle t, \emptyset \rangle \qquad [\![\pi[C]]\!] \Rightarrow [\![\pi]\!].\langle \epsilon, C \rangle$$

where $\pi, \pi_1, \pi_2$ are paths, $t \in T$, and $C \in 2^{\mathcal{L}_Q \cup \mathcal{L}_A}$. That is, the regular expression uses symbols of the alphabet $\Sigma = (T \cup \{\epsilon\}) \times 2^{\mathcal{L}_Q \cup \mathcal{L}_A}$. $\square$

In the following, $FSA(Q)$ will translate a query $Q$ into its associated FSA. This is effected by first computing a regular expression $E$ from $Q$ and then computing an equivalent ($\lambda$-free[6]) nondeterministic FSA for $E$. Furthermore, we will denote by RPQ($\mathcal{Q}$) the inverse operation of FSA($\mathcal{Q}$). Here, we first translate the automaton to a regular expression [15] and then reverse the above transformation to merge labels back into the expression to obtain a query. $\epsilon$-transitions could be introduced more sparingly, but are needed for queries of the form $\cdots (path)^*[cond]$.

**Example 4.5** The query part$^*$[prop[name="engine_id"]] is translated into the regular expression

$$\langle \text{part}, \emptyset \rangle^*.\langle \epsilon, \{\text{prop[name="engine\_id"]}\} \rangle,$$

which is again equivalent to the FSA $\mathcal{Q} = \langle \{q_1, q_2\}, q_1, \delta, \{q_2\} \rangle$ with

$$\delta = \{\langle q_1, \langle part, \emptyset \rangle, q_1 \rangle, \langle q_1, \langle \epsilon, \{\text{prop[name="engine\_id"]}\} \rangle, q_2 \rangle\}$$

Observe that the translation of a query into its regular expression stops at subexpressions (such as prop[name="engine_id"]) that have been moved into a label; those are not further transformed at this point. $\square$

---

[6]A common compositional technique [15] for computing an equivalent FSA for a regular expression needs to introduce dummy transitions which are usually called $\epsilon$-transitions but are not to be confused with ours (thus, we refer to them as $\lambda$-transitions). Our $\epsilon$-transitions which we have introduced to carry query conditions must not be elminated at this point.

**Algorithm** $\langle t_1, C_1 \rangle \circ_{(\mathcal{M}=\langle Q_2, s_2, \delta_2, F_2\rangle, q'_2)} \langle t_2, C_2 \rangle$
**returns** $l \in \langle T \times 2^{\mathcal{L}_Q \cup \mathcal{L}_A}\rangle$ or $\bot$ :
if $(t_1 \neq t_2)$ return $\bot$;
else if$(t_1 = t_2 \neq \epsilon)$ return $\langle t_1, C_1\rangle$;
else if$(t_1 = t_2 = \epsilon)$ {
  $C' := \emptyset$;
  for each $\underline{a = s} \in (C_1 \cap \mathcal{L}_A)$ do {
    if$(\underline{a = s'} \in C_2$ and $s \neq s')$ return $\bot$;
    $C' := C' \cup \{\underline{a = s}\}$;
  }
  for each query $Q' \in (C_1 \cap \mathcal{L}_Q)$ do {
    $\mathcal{M}' := \langle Q_2, q'_2, \delta_2, F_2\rangle$;
    $\mathcal{P}' := \text{FSA}(Q') \bowtie \mathcal{M}'$;
    if$(\mathcal{P}'$ recognizes the empty language) return $\bot$;
    else $C' := C' \cup \{\text{RPQ}(\mathcal{P}')\}$;
  }
  return $\langle \epsilon, C'\rangle$;
}

Figure 5: The $\circ$ algorithm

**Definition 4.6** The binary operation $\mathcal{Q} \bowtie \mathcal{M}$ on automata takes two (nondeterministic) FSA (say, $\mathcal{Q} = \langle Q_1, s_1, \delta_1, F_1\rangle$ and $\mathcal{M} = \langle Q_2, s_2, \delta_2, F_2\rangle$) and computes a new nondeterministic FSA as follows.

$$\mathcal{Q} \bowtie \mathcal{M} = \langle Q_1 \times Q_2, \langle s_1, s_2\rangle, \delta, F_1 \times F_2\rangle$$

with

$$\delta = \{\langle\langle q_1, q_2\rangle, l_1 \circ_{(\mathcal{M}, q'_2)} l_2, \langle q'_1, q'_2\rangle\rangle \mid \langle q_1, l_1, q'_1\rangle \in \delta_1,\ \langle q_2, l_2, q'_2\rangle \in \delta_2,\ l_1 \circ_{(\mathcal{M}, q'_2)} l_2 \neq \bot\}$$

Here, $\circ$ denotes the operation defined by the algorithm of Figure 6.   □

    Note that $\mathcal{A} \bowtie \mathcal{B}$ would compute the product automaton $\mathcal{A} \times \mathcal{B}$ of two nondeterministic FSA if $l_1 \circ_{\ldots} l_2$ would return $l_1$ in case that $l_1 = l_2$ and $\bot$ otherwise. A product automaton $\mathcal{A} \times \mathcal{B}$ recognizes the intersection of the languages recognized by $\mathcal{A}$ and $\mathcal{B}$; as such, product automata are very well suited for restricting a query by the structure of a meta-level database.

    It is easy to see that in case that a query $Q$ does not contain conditions, $\text{FSA}(Q) \bowtie \text{FSA}(M)$ actually is equivalent to $\text{FSA}(Q) \times \text{FSA}(M)$. Otherwise, we check for each condition in a query whether the meta-level database satisfies it. This is the case if

- for attribute assignments: Here, all depends on whether $M$ contains some attribute assignment for the same attribute at the current position. If so, the assignments have to be the same. If such an assignment is missing in $M$, we assume that the restriction applies to instance-level values only, and we go on. [7]

- for path queries $Q'$: We compute an altered meta-level database FSA $\mathcal{M}'$ whose start state is the one the current transition (of which we process the label) leads to. The condition $Q'$ is satisfied by $M$ iff $FSA(Q') \bowtie \mathcal{M}'$ (thus, $\bowtie$ is recursively defined) does not recognize the empty language.

There are various possibilities regarding which restrictions available in $M$ (but not required in $Q$) could be added to the query. For instance, in a semistructured database in which description is made a primitive, we could add the object identifiers of meta-objects as conditions to the query, and by $\mu^{-1}$, we would immediately obtain the state extents of the instance-level objects.

**Theorem 4.7** Let $M$ be a graph database and $Q$ be a query. Then, $\mathrm{RPQ}(FSA(Q) \bowtie FSA(M))$ is equivalent to $Q$ on all graph databases $I$ simulated by $M$. $\square$

**Example 4.8** Let $\mathcal{Q}$ be as defined in Example 4.5 and $\mathcal{M}$ be the automaton for the meta-data of Example 1.1. Then, $\mathcal{Q} \bowtie \mathcal{M} = \langle \{q_1, q_2\} \times \{o_0, \ldots, o_7\}, \langle q_1, o_0 \rangle, \delta, \{q_2\} \times \{o_0, \ldots, o_7\}\rangle$ with

$\delta = \{ \langle \langle q_1, o_0 \rangle, \langle \mathrm{part}, \emptyset \rangle, \langle q_1, o_1 \rangle \rangle,$
$\langle \langle q_1, o_1 \rangle, \langle \mathrm{part}, \emptyset \rangle, \langle q_1, o_2 \rangle \rangle,$
$\langle \langle q_1, o_1 \rangle, \langle \mathrm{part}, \emptyset \rangle, \langle q_1, o_3 \rangle \rangle,$
$\langle \langle q_1, o_1 \rangle, \langle \mathrm{part}, \emptyset \rangle, \langle q_1, o_4 \rangle \rangle,$
$\langle \langle q_1, o_3 \rangle, \langle \epsilon, \{\underline{\mathrm{prop}[\mathrm{name}=\text{``engine\_id''}]}\} \rangle, \langle q_2, o_3 \rangle \rangle \}$

By applying RPQ to $\mathcal{Q} \bowtie \mathcal{M}$, we obtain the pruned query

$$\underline{\mathrm{part.part}[\mathrm{prop}[\mathrm{name}=\text{``engine\_id''}]]}$$

which is equivalent to our original query for all databases described by $M$.

We abbreviate $\underline{\mathrm{prop}[\mathrm{name}=\text{``engine\_id''}]}$ as $X$. To establish, say, the bottom-most of the transitions in $\delta$, we had to evaluate $\langle \epsilon, \{X\} \rangle \circ_{\mathcal{M}, o_3} \langle \epsilon, \{X\} \rangle$ using the algorithm of Figure 6. We obtain

$$\mathcal{Q}' = \langle \{q_1, q_2, q_3\}, q_1, \{\langle q_1, prop, q_2 \rangle, \langle q_2, \langle \epsilon, \{\underline{[\mathrm{name}=\text{``engine\_id''}]}\} \rangle, q_3 \rangle\}, \{q_3\} \rangle$$

and $\mathcal{M}'$ is $\mathcal{M}$ with start state $o_3$. Of course, $RPQ(\mathcal{Q}' \bowtie \mathcal{M}') = X$ and thus $\langle \epsilon, \{X\} \rangle \circ_{\mathcal{M}, o_3} \langle \epsilon, \{X\} \rangle = \langle \epsilon, \{X\} \rangle$. $\square$

---

[7] Using some kind of schema – such as an XML DTD – for $M$, such guesses could be avoided, but here we assume $M$ to be schema-less.

While the schema of Figure 1 needed to be cyclic (through the *part'* association) to be able to represent (in our running example: aggregation) trees, the tree meta-data are acyclic. Although obvious, it is enjoyable to observe that consequently, all pruned path queries are certain to be star-free.

## 5 Conclusions

A main goal of this paper was to propose *description* as a data model primitive, to justify this by examples and likely prospects, and to show it applicable to a wide range of data models beyond the semistructured. In fact, our notion of description is used in many practical large systems today, as it aids in their creation and maintenance. By including description into data model semantics, we may support a wide range of optimizations in an elegant way.

Clearly, our semantics are first-order and can certainly be handled by the most general of previous approaches to semantic query optimization (see e.g. [8]). However, these solve a problem too general to be efficiently manageable in practice.

In the case of semistructured data, our description semantics extends graph schemata by *values* on the meta-level. While this may appear as a minor extension at first, we feel that it allows for a whole range of new applications. Indeed, our meta-data are likely to be human-designed, shared artifacts much like, say, XML DTD's, while graph schemata traditionally have been conceived as automatically generated data hidden to humans. Here we also envision interesting future work.

We have implemented prototype optimizers for both the case of conjunctive queries in an object-oriented data model and semistructured path queries. Experiments risk to be misleading and are not reported in this paper. Clearly, since we have the design authority over both (meta-level) schemata and description meta-data, we can synthesize virtually any optimization speedup we like.

A demo version of the latter system, XDES, a Java framework for optimizing path queries, can be accessed at

http://www.dbai.tuwien.ac.at/staff/koch/xdes/.

This framework is compatible with standard DOM parsers for reading meta-data from XML. Queries to be optimized can be specified in one of several languages, including a large fragment of XPath [33] (here, we are compatible with the Apache Xalan framework) and the language discussed in Section 4 of this paper. We plan to support further languages such as XQuery [34] in the future.